\documentclass[preprint2]{aastex}

\usepackage{graphicx,rotate}
\usepackage{natbib,lscape, amsmath}
\usepackage{latexsym,amssymb,verbatim}
\usepackage{multirow}

\shorttitle{CS lines in hot cores} \shortauthors{Bayet et al.}

\begin{document}


\title{CS lines profiles in hot cores}


\author{E. Bayet\altaffilmark{1}, J. Yates\altaffilmark{1}
and S. Viti\altaffilmark{1}}

\email{eb@star.ucl.ac.uk}


\altaffiltext{1}{Department of Physics and Astronomy, University
College London, Gower Street, London WC1E 6BT, UK.}


\begin{abstract}

We present a theoretical study of CS line profiles in archetypal
hot cores. We provide estimates of line fluxes from the CS(1-0) to
the CS(15-14) transitions and present the temporal variation of
these fluxes. We find that \textit{i)} the CS(1-0) transition is a
better tracer of the Envelope of the hot core whereas the higher-J
CS lines trace the ultra-compact core; \textit{ii)} the peak
temperature of the CS transitions is a good indicator of the
temperature inside the hot core; \textit{iii)} in the Envelope,
the older the hot core the stronger the self-absorption of CS;
\textit{iv)} the fractional abundance of CS is highest in the
innermost parts of the ultra-compact core, confirming the CS
molecule as one of the best tracers of very dense gas.

\end{abstract}






\section{Introduction}\label{sec:intro}

Understanding how stars form at various redshifts is crucial in
order to infer how larger structures such as galaxies are made and
evolve in the Universe. To understand the process of
star-formation, it is essential to determine the properties of the
gas in star-forming regions (hereafter called SFR), both in our
own Galaxy and in external galaxies.

SFR encompass a large range of physical and chemical conditions.
Within SFR, the gas and dust are recycled from prestellar cores to
hot corinos and hot cores. During the star formation cycle, the
pressure, density, temperature and chemistry vary.

Numerous papers present atomic and molecular observational surveys
of prestellar cores, hot corinos and cores in our own Galaxy (e.g.
\citealt{Unge87, Macd96, Prat97, Remi03, Kaif04, Ceca05, Bott07,
Olof07}). Determining the gas and dust temperatures, gas density,
molecular abundances, etc. of each component in SFR can only be
achieved by a close comparison between observations and detailed
modelling.

In this paper, we present a theoretical study of the
$^{12}$C$^{32}$S (hereafter CS) molecular emission from
\textit{hot cores}, motivated by the work of \citet{Doty97} and
\citet{Mill98}. Subsequent papers will present results for molecules
such as methanol, HCO$^{+}$ and HCN. Our first aim is to provide observers with
theoretical CS profiles that may help them interpreting the CS
line emissions arising from hot cores, such as those from
\citet{Wu10}. We do not specifically model any particular hot
core. Instead, we model an archetypical hot core composed of
ultra-compact core and surrounding envelope. We present estimates
of line fluxes and line profiles for comparison with observations.

The study performed by \citet{Mill98} of the hot cores
distinguished two zones of emission: the ultra-compact core
(hereafter UCC) and the Envelope of the hot core. While the UCC
zone is characterized by a size of about 0.03 pc, an average age
of 3.2$\times 10^{3}$ yrs (see \citealt{Mill98}) and an average
density of 1$\times 10^{7}$ cm$^{-3}$, the Envelope corresponds to
a more extended region (0.15 pc size) at a lower density (1$\times
10^{6}$cm$^{-3}$, see again \citealt{Mill98}). In this paper,
studies of both the UCC and the Envelope CS line emissions are
performed between 1$\times 10^{3}$ yrs and 1$\times 10^{5}$ yrs.
It is expected that the Envelope survives longer than the UCC once
the protostar(s) is (are) formed. Thus, its emissions should
remain detectable at later times than the emissions coming from
the UCC. This is why, for the Envelope, we have investigated time
up to 1$\times 10^{6}$ yrs.

The key questions we aim to answer in this paper are \emph{i)}
what are the contributions of these two emitting zones to the
total CS line emissions detected in hot cores? \emph{ii)} How do
these contributions evolve with the age of the hot core? Answering
this question is crucial for improving our current understanding
of massive star formation. CS is particularly useful for
observers, as it emits quite strongly, not only in hot cores in
our own Galaxy (e.g. \citealt{Beut02,Leur07, Wu10}) but also in
external galaxies (see e.g.
\citealt{Mart06a,Baye08b,Alad09,Baye09c}). It is also recognized
as to be one of the best tracers of very dense and warm gas with
line critical densities of about 10$^{6-7}$ cm$^{-3}$
\citep{Plum92, Link80, Snel84}. In addition, its spectroscopic
characteristics are very well known\footnote{See LAMDA:
http://www.strw.leidenuniv.nl/$\sim$moldata/ or BASECOL:
http://basecol.obspm.fr/}.

In Sect. \ref{sec:mod}, we describe the models we use and present,
for the first time, an interface code we have built between the
UCL\footnote{University College London} chemical model (hereafter
called UCL\_Chem) and the radiative transfer code SMMOL. In Sect.
\ref{sec:param}, we specify the parameters used for this
particular study of the CS molecule emission in hot cores. In
Sect. \ref{sec:resu} we present our results and we show how the CS
line profiles and fluxes vary with evolution of the hot core and
different parameters. We discuss the results in Sect.
\ref{sec:disc} and conclude in Sect. \ref{sec:con}.

\section{Model descriptions}\label{sec:mod}

To generate the CS line profiles, we have developed an intuitive
and friendly interface able to couple the UCL\_Chem model and the
radiative transfer code SMMOL. This Interface code will be very
shortly publicly available. The UCL\_Chem model is briefly
described in Subsect. \ref{subsec:chem} whereas a summary of the
main characteristics of SMMOL is presented in Subsect.
\ref{subsec:SMMOL}. The interface is described in Subsect.
\ref{subsec:Inter}.

\subsection{The UCL\_Chem model}\label{subsec:chem}

The UCL\_Chem model is fully described in \citet{Viti99} and its
upgrades are presented in \citet{Viti04} and \citet{Baye08a}.

The UCL\_Chem is a time-dependent 1-D chemical code that can be
used to model the evolution of the gas and dust during the
formation of a star. Here we used it to simulate the formation of
a hot core. As in \citet{Viti04}, we first model the collapse of a
10K core (Phase I); we then follow the chemical evolution of the
remnant core once the star is born (Phase II). The presence of an
infrared source in the centre or in the vicinity of the core is
simulated by subjecting the core to an increase in the gas and
dust temperature.

In both phases, the chemical network is based on more than 1700
chemical reactions taken from the UMIST database
\citep{Mill97,LeTe00} involving about 200 species, of which 42 are
surface species. The relevant surface reactions included in this
model are assumed to be only hydrogenation, allowing chemical
saturation where this is possible.

One of the outputs of the UCL\_Chem is the fractional abundance
(with respect to the total number of hydrogen nuclei) of gas and
surface species. See Sect. \ref{sec:param} for a description of
the grid of UCL\_Chem models ran for this study.

\subsection{The SMMOL model}\label{subsec:SMMOL}

The molecular line radiative transfer code solves the multilevel
radiative transfer problem in a 1-d spherical geometry. The code
we use, SMMOL, is based upon two codes; Multi-Mol \citep{Yate97}
and the SMULTI code developed by \citet{Harp94}. SMMOL uses an
Accelerated Lambda Iteration (ALI) scheme to speed the convergence
of the iterative scheme that is used to solve a set of linearly
perturbed kinetic master equations, in order to determine  the
steady state populations of a molecule's energy levels and the
radiation field.  This is the MULTI method described in
\citet{Schar85}. Subsequently the MALI method \citep{Humm92} was
added to SMMOL; this uses an ALI technique to speed the
convergence of a set of preconditioned kinetic master equations.

SMMOL (Spherical Multi-MOL) is a general non-LTE molecular line
radiative transfer code that has reproduced the spectral lines
observed towards a large number of sources (e.g. \citealt{Bene06,
Yao06, Lera08, Tsam08, Lera10}). It is fully described in
\citet{Rawl01} and has been benchmarked with other radiative
transfer codes in \citet{VanZ02}. We recommend these papers to any
reader. The output line profiles are convolved with user supplied
telescope properties using the method described in \citet{Scho88}.

Typically SMMOL uses 400 rays to compute the spherical cloud
modelled to compute the intensities at each radial grid-point
(hereafter "shells") in the cloud. The code is capable of adapting
its sampling along each ray to take  account of large velocity
changes between shells e.g. if the line of sight velocity change
between adjacent shells causes the individual line absorption
profiles at these adjacent shells to be non overlapping in
frequency space; this is not physical and can allow photons to
escape the cloud that would otherwise have been absorbed.

\subsection{The Interface code}\label{subsec:Inter}

The Interface is a fortran 95 programme, that transforms the
output from UCL\_Chem into line fluxes and profiles via the use of
SMMOL. It is able to model various gas phases from diffuse gas to
hot cores. It is currently being developed for AGB stars and
planetary nebulae.

The output file of the UCL\_Chem model (i.e. fractional abundances
as a function of time and depth) is the input file of the
Interface. The UCL\_Chem provides a grid in optical depth (Av) of
fractional abundances of about 200 species, at various time steps.
At each time step, the grid in Av has to be adapted to the spatial
(linear) grid of shells used later on in SMMOL and described by:
\begin{equation}\label{eq:shell}
r_{i}=\frac{Av\times(1.6\times10^{21}/met)}{density} - (\Delta
r/2.0+(\Delta r\times i))
\end{equation}
where $met$ is the metallicity assumed in the UCL\_Chem and
$\Delta r$ is the thickness of the shells (assumed all equal here
- linear distribution of the shells):
\begin{equation}\label{eq:thick}
\Delta r = \frac{size}{n_{shell}}
\end{equation}

Then, the Interface also manages the allocation, to each shell of
the appropriate physical values, required to run SMMOL. They are
the density, temperature, fractional abundance etc which are
extracted from UCL\_Chem. To do so, the Interface converts Av into
distance and interpolate the UCL\_Chem values alongside the shell
grid using SPLINE and SPLINT functions\footnote{Both functions are
Numerical Recipes routines}. For this study, the dust temperature
is assumed to be equal to the gas temperature since in the
UCL\_Chem model, the dust temperature is not calculated. For our
models (see Sect. \ref{sec:param}), this assumption is valid since
the opacity and the density are very high in both the
Ultra-compact core and the Envelope (Av $\gg$ 100 mag). Once the
Interface has created the correct grid, it automatically runs the
SMMOL programme. The outputs of SMMOL are plotted and tabulated by
the Interface.

We hope that the Interface will be used to automatically interpret
data in from space telescopes (e.g. Herschel, JWST) and
observatories (e.g. ALMA, e-Merlin...).

\section{Choice of parameters}\label{sec:param}

\subsection{General assumptions}\label{subsec:gene}

The UCL\_Chem models are converted into inputs for the SMMOL code
using the Interface code, as described in Sect.
\ref{subsec:Inter}. We describe here our choice of parameters and
assumptions.

The main sampling parameters such as the number of radial density
shells, and the line-of-sight frequency sampling through the
cloud, are determined beforehand to ensure that the output fluxes
are invariant with respect to sampling for all the models run
through SMMOL. We found that $n_{shell}=100$ and 400
lines-of-sight for ray tracing, ensured that the results were
invariant; the smallest clouds we sample are actually very
oversampled.

We used the first 40 rotational levels of CS in the vibrational
ground state; the molecular data and the collisional rates with
respect to to H$_{2}$ are from the Cologne Database for Molecular
Spectroscopy\footnote{See the CDMS website:
http://www.astro.univ-koeln.de/cdms/}.

The kinetic temperature law (see Eq. \ref{eq:temp}) we used is a
compromise between the need for a power law which is flatter in
the inner hot core (to take into account photon trapping effects
due to higher optical depths, which slow the cooling of material
by radiation) and the need for a 1/r$^{0.5}$ law to describe the
cooling we would see in the lower optical depth outer cloud. We
have used Eq. \ref{eq:temp} to be consistent with previous studies
\citep{Mill98,Viti99,Bene06,Lera08,Lera10}.

We chose a turbulence velocity of 1.5 kms$^{-1}$ (since
observations show narrow line profile as in \citealt{Hatc98b})
assumed no velocity gradient, a typical distance of 450pc
(Orion-KL: see \citealt{DeVi02}) to the source and solar
metallicity. If these hot cores had a large velocity gradient, it
would reduce the optical depth of the lines and allow photons to
travel more freely in the hot core. The spectra will therefore
become less absorbed and line broadening due to optical thickness
would be reduced. However the line width would be increased as
emission would come from a greater range of velocities. Actually,
there is currently little evidence for velocity gradients in these
systems. Observations of lines from these systems show narrow
lines (e.g. \citealt{Hatc98b}) suggesting that all parts of the
UCC and Envelope can be radiatively coupled. The line absorption
profile is clearly dominated by turbulence, with the kinetic
temperature contributing to the FWHM.

The cloud was illuminated by the standard interstellar radiation
field \citep{Habi68} and the dust extinction model is from
\citet{Math90}. The emergent spectra were convolved with the
appropriate telescope beam (either JCMT or IRAM). To enable us to
make useful predictions for ALMA or any interferometric
observations (e.g. CARMA, PdBI,...), we need to predict emission
from the source at sub-arcsec resolution. The 24 models whose
parameters are displayed in Table \ref{tab:models} have been run
twice, once with IRAM/JCMT resolutions (line profiles presented as
such, see Fig. \ref{fig:2}-\ref{fig:4}), and once with effectively
an infinite resolution (radial distribution, see Fig. \ref{fig:5})
allowing us thus to get the largest range of (sub-arcsec)
resolutions possible. The gas density (see Eq. \ref{eq:dens}),
fractional abundances and cloud size are provided by the UCL\_Chem
model (see Fig. \ref{fig:1}).

Currently the high resolution observational data that exist
suggest that objects are ellipsoidal or spherical in shape
\citep{Davi10,Grav10,Wu10}. We are constrained by the 1-D nature
of the UCL\_Chem and SMMOL models, which means the only models we
can construct are spherical. The most likely effect of
non-spherical geometry would be the scenario where collapse was
aided by magnetic field lines, giving a flattened density profile
with an equatorial enhancement of material. The inclination of the
source now becomes important; a pole-on source will have lower
optical depth than an equatorial-on source. This means we could,
for instance, overestimate the number of low optical systems, and
underestimate the mass of objects.

We ran the Interface code for several ages from $1\times 10^{3}$
yrs to $1\times 10^{6}$ yrs (see Table \ref{tab:models}).
\citet{Mill98} assumed more specifically a typical age of
$3.2\times 10^{3}$ yrs for the Ultra-Compact Core (see Models
HC$_{1}$, HC$_{3}$, HC$_{5}$ and HC$_{7}$ to HC$_{13}$) whereas a
longer time for a less dense gas such as the one contained in the
Envelope (see Models HC$_{2}$, HC$_{4}$, HC$_{6}$ and HC$_{14}$ to
HC$_{24}$) is expected.

\subsection{UCC and Envelope specific assumptions}\label{subsec:spec}

To reproduce the CS line emission in hot cores for various density
and temperature structures and ages, we have run over 80 UCL\_Chem
models\footnote{The grid of models can be found at
    http://www.homepages.ucl.ac.uk/$\sim$ucapdwi/interface/. Still under
    construction, this website aims at providing a variety of
    UCL\_Chem  models that the user can couple with the radiative transfer code SMMOL
    to obtain, for a various range of physical and chemical conditions, the line
    intensities and line profiles of more than 200 species.} but present here only results from the most interesting
ones; the parameter choices made for these 24 calculations are
summarized in Table \ref{tab:models}.

The size of the region has been set to 0.03 pc for the
Ultra-Compact Core and 0.15 pc for the Envelope. Other input
parameters such as the FUV radiation field, the cosmic ray
ionisation rate, the gas-to-dust mass ratio,... are all set to
their standard values as in \citet{Baye08a} (see Tables 1, 2 and 3
in their paper).

To represent the UCC zone, we ran the UCL\_Chem model with a
temperature of 10 K in Phase I, collapsing the core to a critical
density of 10$^{7}$cm$^{-3}$ (central density: $n_{c}$). We chose
a central density of $10^{7}$cm$^{-3}$ in order to be consistent
with our previous work \citet{Baye08a} and because it is the
density derived by single-dish observations \citep{Hatc98a}. In
fact, the central density of hot cores may be higher than that
(see interferometric data in \citealt{Belt05}). However due to the
high opacity of hot cores (Av $\gg$100 mags) a small change in the
central density should have a negligible effect on the fluxes.

In both regimes (i.e. UCC and Envelope), when no temperature and
density profiles are applied (see Models HC$_{1}$ and HC$_{2}$),
the temperature and the density are kept fixed at 300 K and $1.0
\times 10^{7}$ cm$^{-3}$, respectively, for the UCC zone, and to
101.5 K and $3.8 \times 10^{6}$ cm$^{-3}$, respectively, for the
Envelope case. For Models HC$_{3}$ and HC$_{4}$, we only applied a
density profile (see Sect. \ref{subsec:stru}).

\begin{figure*}
\hspace*{0.5cm}\includegraphics[height=8cm]{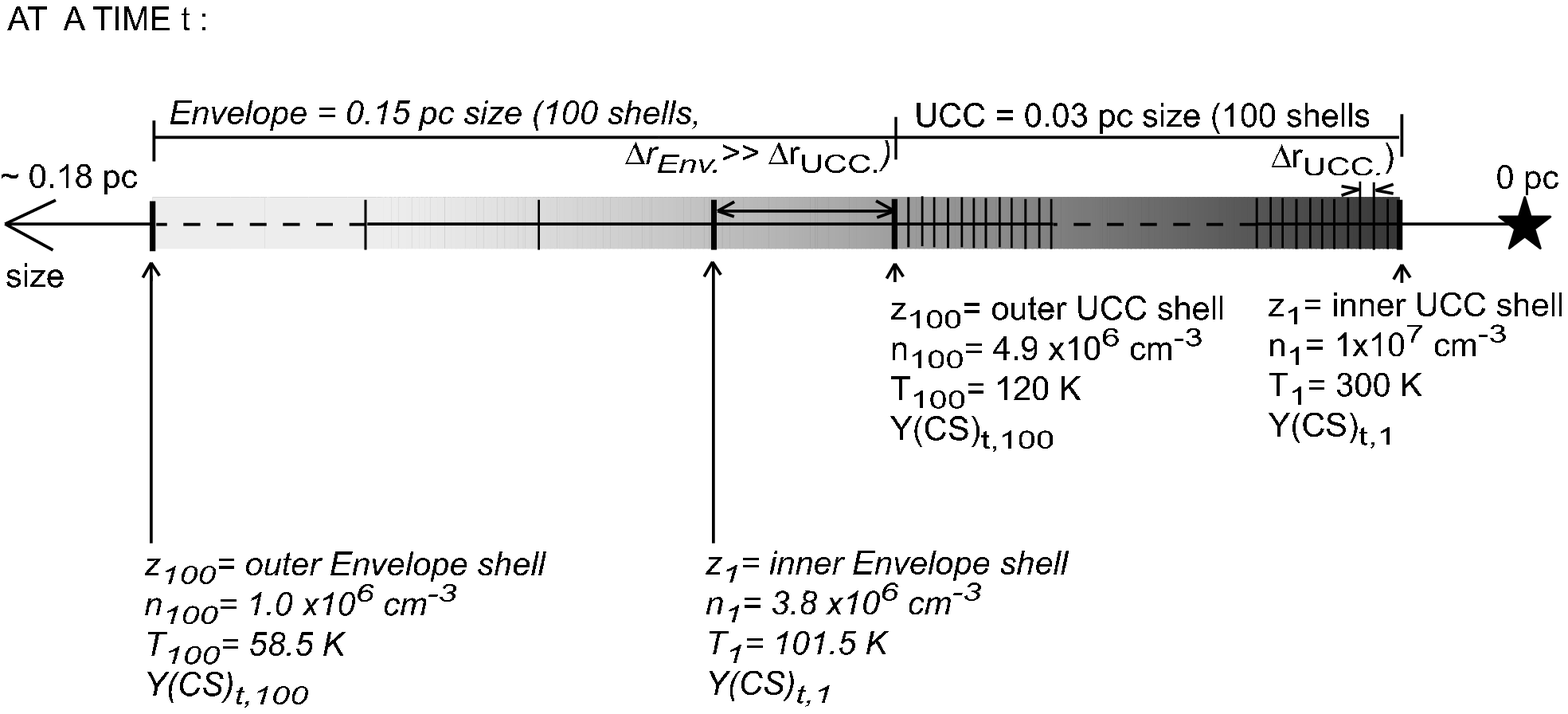}
\caption{Cartoon representing the structure of the hot core we
adopt: the Ultra-compact core UCC (right hand side) and the
Envelope (left hand side). The characteristics of each zone, in
the multi-points approach, for a certain time t, are specified.
The grey scale does not have any physical meaning but should be
considered more as an guideline for the eyes. It symbolises the
gradient in density and temperature through the hot core with the
darkest zones corresponding to the highest values of density and
temperature. The shells of width $\Delta r$ are represented by
vertical bars from number 1 to 100 (z$_{1}$ showing the location
of the shell number 1). Regular and italic fonts are used for
distinguishing the UCC shells from the shells of the Envelope,
respectively.}\label{fig:1}
\end{figure*}

When a temperature profile is applied (this was done for Models
HC$_{5}$ and HC$_{7}$ to HC$_{13}$ using Eq. \ref{eq:temp}), the
temperature varies from 120 K to 300 K (between the lowest to the
highest Av, respectively). We used the formula seen in
\citet{Viti99}:
\begin{equation}\label{eq:temp}
T(r)= T_{c} \times (\frac{r}{r_{0}})^{-0.4}
\end{equation}
where $T_{c}$=300K is the central temperature typical for hot
cores (see \citealt{Mill98, Viti04}) and $r_{0}$=0.18 pc is the
distance from the edge of the hot core to the newly born star (see
Fig. \ref{fig:1}). When a density profile is applied (e.g. Models
HC$_{3}$, HC$_{5}$ and HC$_{7}$ to HC$_{13}$ and Eq.
\ref{eq:dens}) the density profile, from the lowest to the highest
Av, leads to the density varying from $4.9 \times 10^{6}$
cm$^{-3}$ to $1.0 \times 10^{7}$ cm$^{-3}$, respectively. We used
the same formalism as in \citet{Case95, Hatc00} and
\citet{Bacm00}, which is effectively a Bonnor-Ebert sphere
approximation:
\begin{equation}\label{eq:dens}
n(r)= n_{c} \times (1+\frac{r}{r_{c}})^{-1.5}
\end{equation}
where $n_{c}=10^{7}$cm$^{-3}$ is the density assumed at the center
of the hot core and $r_{c}=0.05$ pc \citep{Nomu04} is the radius
between isothermal and non-thermal velocity effects in hot cores.

To model the Envelope, we have run the UCL\_Chem model with a
temperature of 10 K in Phase I, letting the UCC collapse to a
critical density of $3.8 \times 10^{6}$cm$^{-3}$ (i.e. the density
of the inner shell of the Envelope, see Fig. \ref{fig:1}). From
the lowest to the highest Av, when a temperature profile is
applied (e.g. Models HC$_{6}$ and HC$_{14}$ to HC$_{24}$ using Eq.
\ref{eq:temp}), the temperature varies from 58.5 K to 101.5 K,
respectively. When a density profile is applied (e.g. Models
HC$_{4}$, HC$_{6}$ and HC$_{14}$ to HC$_{24}$ using Eq.
\ref{eq:dens}) the density varies from $1.0 \times 10^{6}$
cm$^{-3}$ to $3.8 \times 10^{6}$ cm$^{-3}$, respectively.

\section{Results}\label{sec:resu}

In the following section, we present the results for our study to
note the effects of the variations in the internal structure of
the hot core (see Subsect. \ref{subsec:stru}), and its age (see
Subsect. \ref{subsec:age}), and we compare the spectra from the
UCC and the Envelope (see Subsect. \ref{subsec:zone}), having
addressed velocity and geometry effects in Sect.
\ref{subsec:gene}. Figures \ref{fig:2}-\ref{fig:4} show examples
of the most interesting changes affecting the line profiles
whereas Table \ref{tab:models} summarizes for each hot core model
the integrated line fluxes of the CS transitions obtained.

\subsection{Influence of the density and temperature profiles on
the CS line emissions}\label{subsec:stru}

\begin{figure*}
\hspace*{1cm}\includegraphics[height=15cm]{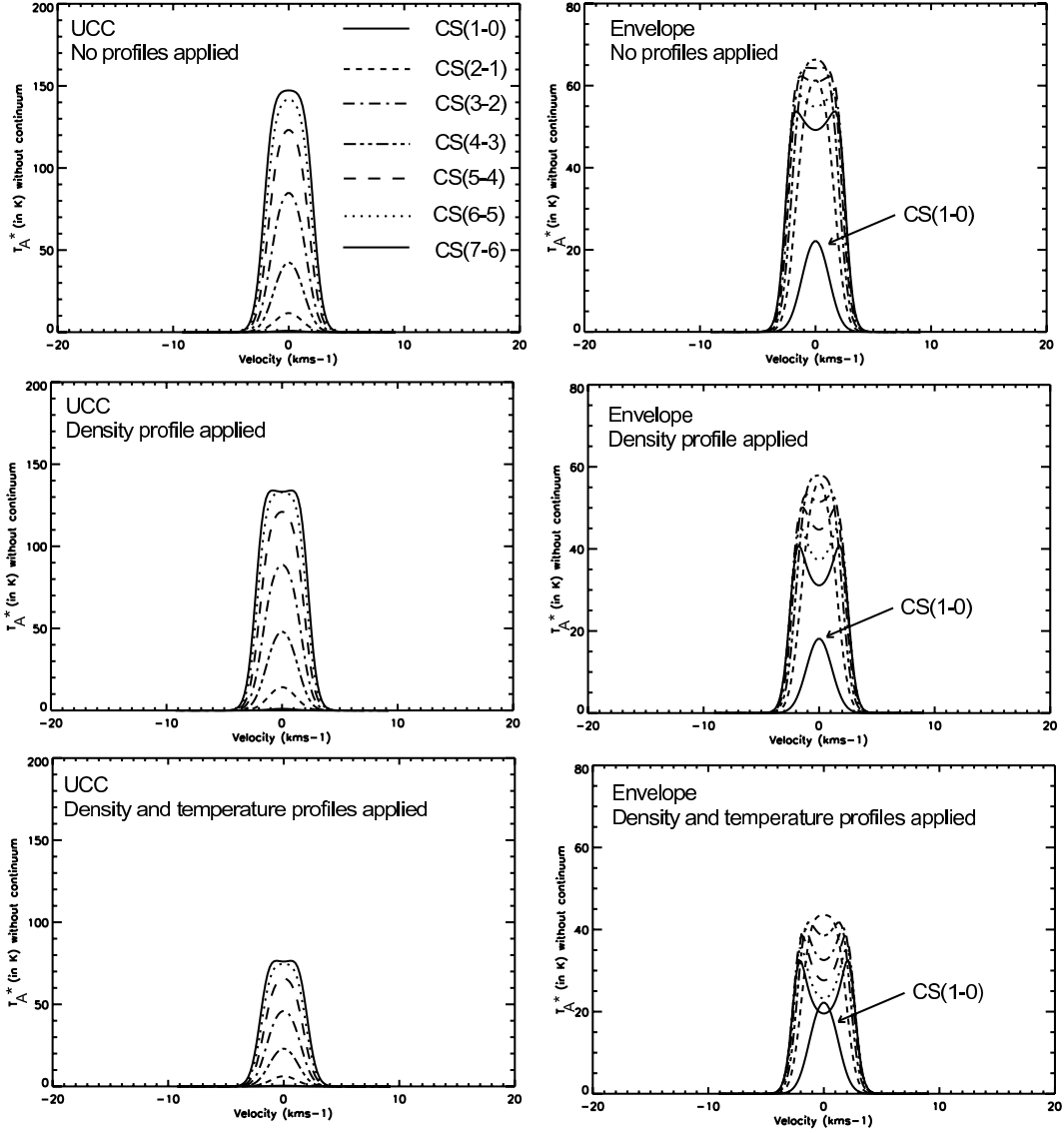}
\caption{Variation of the CS line profiles for the UCC (left, from
top to bottom: Models HC$_{1}$, HC$_{3}$ and HC$_{5}$) and the
Envelope (right, from top to bottom: Models HC$_{2}$, HC$_{4}$ and
HC$_{6}$) when various internal hot core structures are used. From
top to bottom, we have applied no profiles, only a profile in
density, and a profile in both density and temperatures (see
Sect.\ref{sec:param} and Subsect.\ref{subsec:stru}). The CS line
profiles are expressed in antennae temperature (in K) using the
IRAM and the JCMT resolutions for low-J and high-J CS lines,
respectively. }\label{fig:2}
\end{figure*}

\begin{figure*}
\rotatebox{90}{\includegraphics[height=16cm]{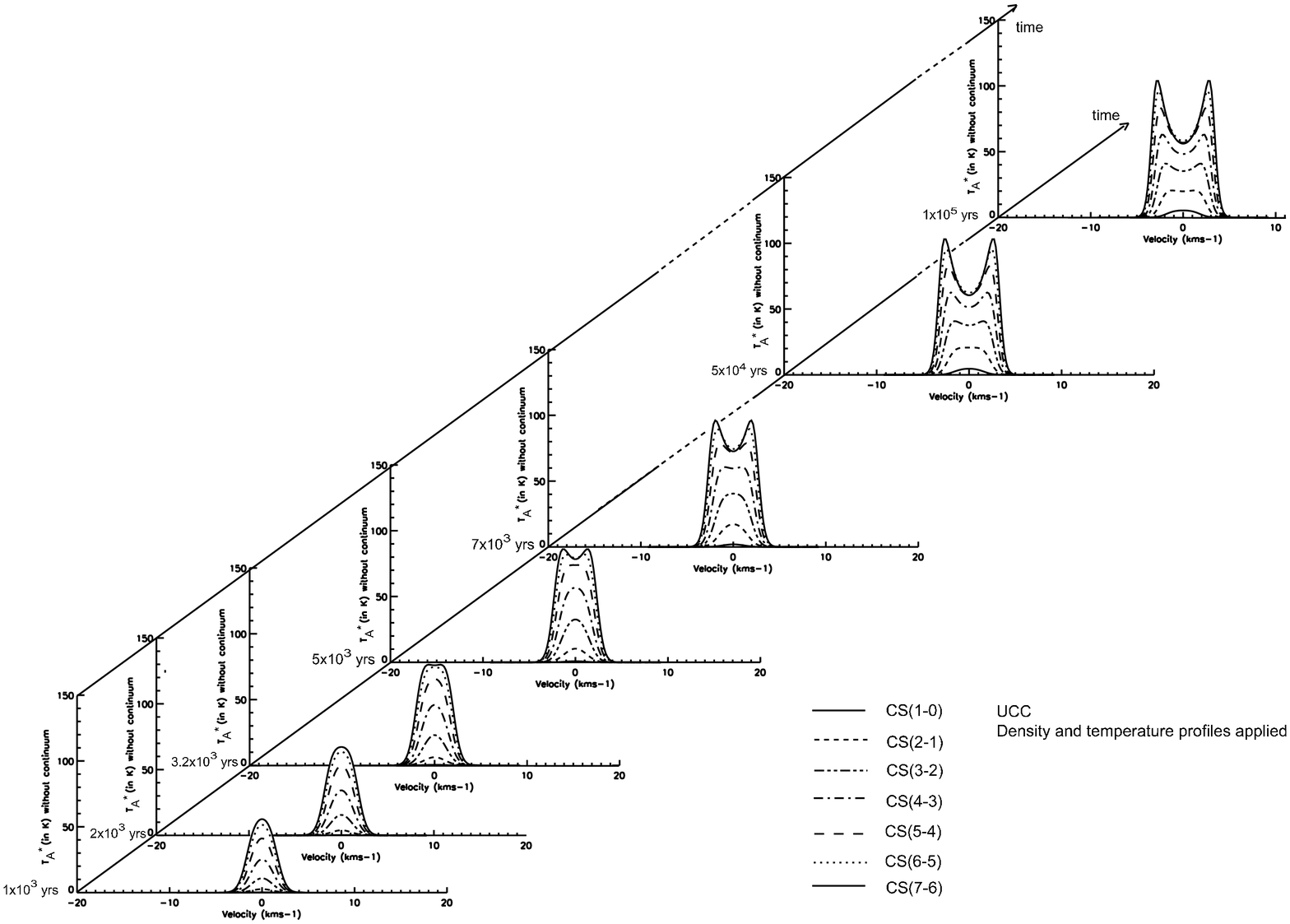}}
\caption{Variation of the CS line profiles for the UCC with time
(age plotted in the third dimension: Models HC$_{7}$, HC$_{8}$,
HC$_{5}$, HC$_{9}$, HC$_{10}$, HC$_{12}$ and HC$_{13}$,
respectively).}\label{fig:3}
\end{figure*}

\begin{figure*}
\rotatebox{90}{\includegraphics[height=16.5cm]{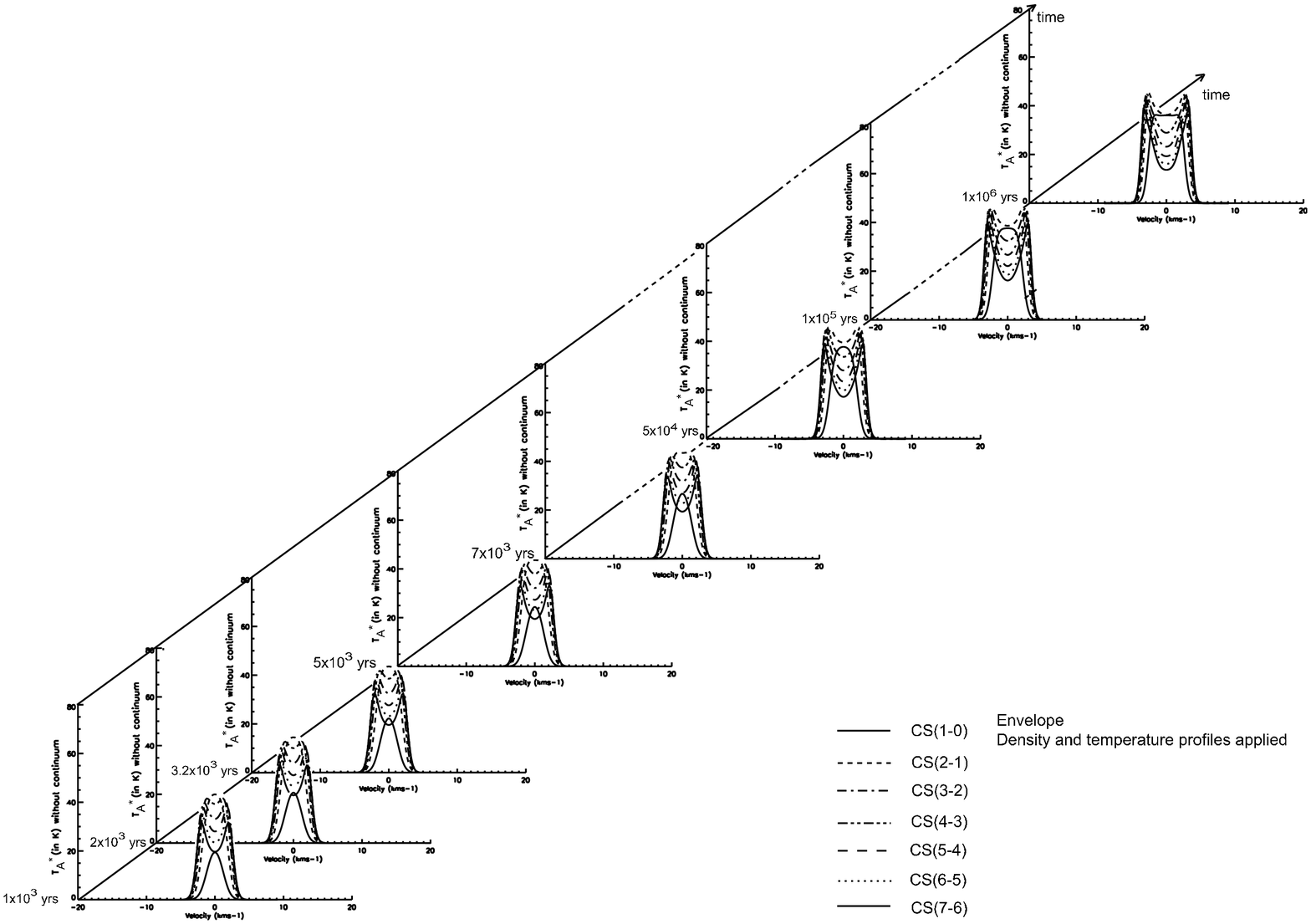}}
\caption{Variation of the CS line profiles for the Envelope with
time (age plotted in the third dimension: Models HC$_{14}$,
HC$_{15}$, HC$_{6}$, HC$_{16}$, HC$_{17}$, HC$_{20}$, HC$_{22}$
and HC$_{24}$, respectively). Since it is expected that the
Envelope survives longer than the UCC zone, our study has been
extended up to $1 \times 10^{6}$ yrs.}\label{fig:4}
\end{figure*}

The influence of the density and temperature profiles on the CS
line emissions \footnote{derived from the models} is shown in Fig.\ref{fig:2} (from bottom to top).
This figure represents thus results obtained for Models HC$_{1}$,
HC$_{3}$ and HC$_{5}$ (for the UCC) and Models HC$_{2}$, HC$_{4}$
and HC$_{6}$ (for the Envelope) whose parameters are seen in Table
\ref{tab:models}.

In the case of constant density (top plots of Fig. \ref{fig:2}),
it is interesting to note the differences in the line profile
shapes of the low-J CS lines (up to CS(3-2)) as compared to those
of the higher-J CS transitions (from CS(4-3)). Firstly, high-J CS
lines show the strongest emissions. Secondly, the low-J CS lines
have a narrower line width than the high-J CS transitions (by a
factor of about $\lesssim$ 1.5-2.0). The first result can be
understood by looking at the level population distribution.
Indeed, in Model HC$_{1}$, the majority of the collisions occur in
the high levels of CS, favoring transitions at high-J rather than
low-J, whose levels are less populated by one order of magnitude
on average. In addition, the transitions between the high-J levels
have high $A_{ij}$ coefficients ($A_{ij}$ is proportional to
$\nu^{3}$). These factors give also rise to higher source function
terms and so to brighter emission. In parallel, the high-J lines,
as well as being brighter than low-J transitions, are also broader
and show more spectral structure than the low-J transitions.
Indeed, the higher-J lines have flatter peaks and some show strong
self absorption at the systemic velocity. The line broadening is a
consequence of opacity, more often seen in stellar spectra and
often called the curve-of-growth. The low optical depth lines have
narrow Gaussian line shapes. As the optical depth increases the
line centre emission saturates, however the line wing emission can
still increase and it is this increase that broadens the line and
increases the FWHM of the line. Eventually the line wings saturate
and flattened line-shapes are observed. Finally the optical depth
at line centre is high enough to promote self absorption and
twin-peaked spectra are formed. The observed molecular line
curve-of-growth spectral behavior can give a spectral signature
for warm dense hot cores.

To disentangle the influence of the density and the temperature
profiles on the CS line emissions, we have first kept the
temperature constant to 300 K in the UCC and 101.5 K in the
Envelope whatever the UCC (Envelope) radial shell (see Models
HC$_{3}$ and HC$_{4}$, respectively). In such case (density
profile only), the UCC CS line profile shapes do not seem
significantly changed as compared to the case where there is no
profile. On the contrary, for the Envelope, where the difference
in densities between the outer and the inner shells is steeper
than for the UCC, we see more significant changes.

However, one notes that for Model HC$_{3}$, a slight broadening of
all the lines is seen, between a factor of 0.33 and 0.16 for the
CS(1-0) and the CS(7-6) transitions, respectively. In addition,
the peak antennae temperatures are on average weaker than in the
case where no profile is applied (differences varying between 3 K
and 12 K). A saturation in the CS(7-6) profile is also seen. This
is due to the higher fractional abundances of CS obtained from the
chemical model (see Table \ref{tab:models}).

Finally when we add the temperature structure (see Model HC$_{5}$
for the UCC and Model HC$_{6}$ for the Envelope), we see that
(Fig. 2), the peak antennae temperatures of all the CS lines but
CS(1-0) in the Envelope are indeed weaker by a factor of 1.5.
To be more precise, we found that the modelled Tpeaks
change in all 80 models when we implement a $r^{-0.4}$ temperature
profile as compared to their values without. It means that the
distribution of the temperatures inside the hot core, i.e. the
temperature variation seen from shell to shell, do have a
consequence for the integrated ("total") profiles of CS lines. The
differences in the Tpeak values range from 20\% to a factor of
two, depending on the line and the source considered (see in Fig.
2 the bottom two plots for UCC and Envelope: both show a decrease
of the CS line Tpeak when a temperature profile is implemented).
We believe that ALMA may be sensitive to factors as small as this
in Tpeak. In fact already with the CARMA and IRAM-Plateau de Bure
Interferometer such factors are detectable for hot cores
\citep{Wu10}. It may therefore be possible to estimate Tc and
potentially reconstruct the temperature profile of the observed
source by using Eq. 3.

All the lines (except J=1-0) are thus good probes of these changes in
temperature. The peak antennae temperature is a measure of the gas
kinetic temperature if the gas is in LTE, the column is optically
thick and the source is resolved by the telescope at the observing
frequency.

The CS(1-0) does not seem affected at early times by the changes
in temperature in the Envelope. This might come from the fact that
the fractional abundance of CS is quite low in the Envelope. The
$A_{ij}$ is also 300 times less for the J=1-0 transition compared
to the J=7-6 transition. Also the energy of the J=1 level
$E_{(J=1)}=2.35$ K, which means that in a gas of 50-100 K its
population is comparatively less than in clouds where T=10 K. This
is why the 1-0 line is the weakest line in both the Envelope and
the UCC. However the fractional abundance of CS increases by 20 in
the Envelope during its evolution and that is why the line
strength grows and the line starts to become flat topped.

\subsection{Influence of the age of the hot core on the CS line
emissions}\label{subsec:age}

Figure \ref{fig:3} shows the evolution of UCC CS lineshapes with
the evolution of the hot core. We see the high-J lines increase in
flux with time and become increasingly flattened, broader and self
absorbed; some eventually produce twin peaked spectra.  The low-J
lines grow to large intensities with time and also show flattened
profiles at large times.

Figure \ref{fig:4} shows the evolution of Envelope CS lineshapes
with the evolution of the hot core. It shows that transitions for
J=5, 4, 3, 2 are initially the brightest transitions, with the
J=2-1 transition being the brightest line; the J=3 to 7
transitions are self absorbed at all times, with the J=7-6 line
being the broadest line. The J=1-0 transition is initially the
least bright and narrowest line.

By the largest time, the J=7 to 2 transitions are now all self
absorbed and have basically the same width. There has been a
modest increase in the peak flux of these lines. The J=1-0 line is
now very bright, broader and has a flat top.

The UCC lines are brighter than the Envelope lines. For the
Envelope, the best tracer of age seems to be the CS(1-0)
transition which shows the largest variations in fluxes with
respect to time (see Models HC$_{14}$ to HC$_{24}$). In principle,
this transition could be used as an evolutionary indicator.
However we note that the sulphur-bearing chemistry depends very
critically on the gas temperature and density (as also found by
\citealt{Wake04, Viti04}) and hence, care must be taken to
constrain initial conditions. In the case of the UCC, the
differences in fluxes from $1\times 10^{3}$ yrs to $1\times
10^{6}$ yrs may not be large enough to be detectable.

\subsection{Influence of the size of the hot core on the CS
line emissions}\label{subsec:zone}

There are, as expected, clear differences between UCC CS line
emissions and those coming from the Envelope (see Figs.
\ref{fig:2}, \ref{fig:3} and \ref{fig:4} and Table
\ref{tab:models}). We restrict our comparison to the most
realistic cases which are the cases where both density and
temperature profiles are applied to the two zones.

A first interesting remark concerns the radial distribution of the
CS fractional abundance, for a given time. From column 9 of Table
\ref{tab:models}, for a fixed time (e.g. $3.2\times 10^{3}$ yrs )
one notes that, from the inner UCC shell (closest to the star) to
the outer Envelope shell (edge of the hot core), the CS fractional
abundances does not increase nor decrease constantly (see
  paired models in Table \ref{tab:models}). In fact,
Table \ref{tab:models} shows that the CS fractional abundance is
predominantly produced in the UCC inner zone of the hot core,
confirming CS as preferentially linked with very dense gas, and
therefore could be considered as an ideal tracer of this gas phase
(see also \citealt{Link80, Snel84}). For example, we can
see that for the coupled models HC$_{3}$ and HC$_{4}$, the UCC
(HC$_{3}$) abundance of CS is $1.72 \times 10^{-8}$ whereas the
Envelope (HC$_{4}$) can account for only $3.77 \times 10^{-9}$.
This is a factor of 4.6 difference. In other words, the UCC
produces 4.6 times more CS (in abundance) than the Envelope. For
other models, the difference are of 2.6 (HC$_{8}$ and HC$_{15}$),
7.2 (HC$_{9}$ and HC$_{16}$), 9.6 (HC$_{10}$ and HC$_{17}$), 12.1
HC$_{11}$ and HC$_{18}$), 17.6 (HC$_{12}$ and HC$_{20}$) and 11.5
(HC$_{13}$ and HC$_{22}$). The only models where the CS
is not predominantly produced by the UCC is the pair
HC$_{7}$ and HC$_{14}$ (difference of 0.9 only).

\begin{figure*}
\includegraphics[height=6.5cm]{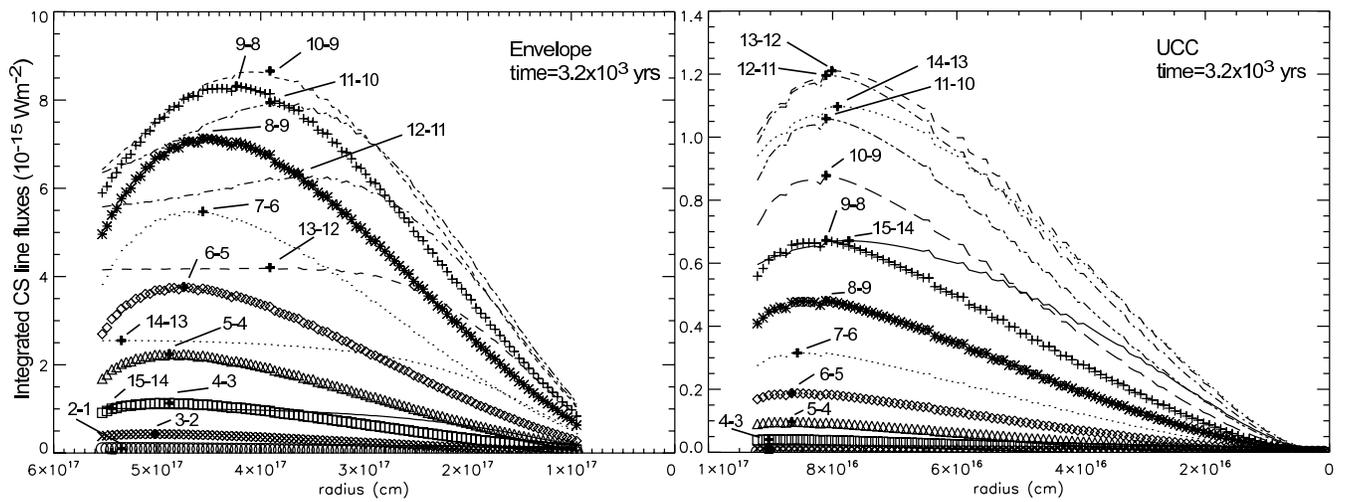}
\caption{Variation of the CS integrated line fluxes for the UCC
(right hand side) and the Envelope (left hand side). See text in
Subsect. \ref{subsec:zone}}\label{fig:5}
\end{figure*}

The radial integrated CS line
fluxes distribution is shown in Fig. \ref{fig:5} at $3.2\times
10^{3}$ yrs (as assumed in \citealt{Mill98}). We note that the
line flux distribution as a function of radius at different times
(not shown) are the same as those shown in Fig. \ref{fig:5}. Note
that, for the plots shown in Fig. \ref{fig:5}, we did not apply
any telescope convolution when we have run SMMOL because we wanted
to identify all the emission from the UCC and the Envelope (hence
the differences in CS line fluxes between Table \ref{tab:models}
and Fig. \ref{fig:5}). For both plots in Fig. \ref{fig:5}, the
protostar is located at the origin of the x-axis (right hand side)
following the convention adopted in Fig \ref{fig:1}. The maxima
integrated CS line fluxes are located on the plots by a thick
black cross. The maxima are all distributed in the UCC within a
restricted radius ($8-9\times 10^{16}$ cm from the source) whereas
for the Envelope, their position is more spread ($3.5-5.5\times
10^{17}$ cm from the source). In the Envelope, the CS lines fluxes
maxima, from 1-0 to 10-9 are distributed deeper and deeper inside
the gas. A turnover occurs for the maximum of the CS(10-9) line
flux. From this transition onwards up to the CS(15-14) line, the
maxima location is moving towards the edge of the Envelope. The
turnover is actually controlled by the competition between the
source function and absorption terms at each radius. These terms
depend upon volume of gas at a radius, the population of the
levels and the Einstein $A$ coefficients of the transitions (Note
that $A_{ij}$ increases with J). In the Envelope this produces
most flux for the J=10-9 transition. The turnover at large radii
is caused by the reduction in the source function and optical
depth due to decrease of gas density and temperature.

In the UCC, the same factors affect the line emission as for the
Envelope. The higher density and temperature explain why the
J=13-12 emission is the brightest transition. The turnover happens
proportionately much closer to the UCC outer edge than in the
Envelope and thus does not appear clearly. This is because the
optical depth and source functions of the optically thick lines
drops due to falling densities and temperatures and these actually
cause the low J lines to turnover as well.

Here, we have simplistically assumed that when a hot core is
observed the total CS flux (taking into account the
self-absorption) is, to a first order approximation, equal to the
sum of the emission coming from the UCC and from the Envelope.
\citet{Mill98} in their study, assume that hot cores can be as
large as 1 pc size and consequently that there is a potential
third contribution to the total molecular emission from the halo
(see \citealt{Mill98}). We did not investigate such region in this
paper since the halo is much more diffuse than the Envelope and
the UCC, and the CS may not be a good tracer of such gas.

As seen from Fig. \ref{fig:5}, contrarily to the CS(1-0), (2-1)
and (3-2) line fluxes, it is interesting to note that the high-J
CS line fluxes (from 6-5) are mainly emitted from the UCC zone
(factors of differences between the Envelope and the UCC
contributions from 3.6 to 13.1 - see also Table \ref{tab:models}).
This makes the low-J and the high-J CS lines better tracers of the
Envelope and the UCC, respectively. This result can only be
confirmed by interferometric observations. The envelope has a
lower hydrogen number density and kinetic temperature, i.e.
$E_{(J=7)}=49.3$ K than the UCC. The CS rotational high-J levels
in the UCC are therefore more highly populated that the same
levels in the envelope. Although the UCC is smaller than the
Envelope the excited column at high-J levels in the UCC is
brighter than that in the Envelope. In the UCC (T=100-300K) the
J=1 level is very underpopulated as compared to the Envelope and
therefore has a much lower optical depth and source function; it
is for this reason that despite the fractional abundance of CS
increasing we see a weak J=1-0 line at all times in the UCC.

Finally we note that in the case of the Envelope, the lines
between J=2-1 and J=6-5 show similar widths (within a factor of
1-2 kms$^{-1}$), contrarily to the UCC case where line widths are
quite different from a transition to another. The high-J lines
remain optically thin for much longer in the UCC than for the
Envelope. The absorption profile of each line is slightly broader
because of the higher kinetic temperature in the UCC (0.2-0.3
kms$^{-1}$)

\section{Discussion}\label{sec:disc}

Unfortunately, there are no complete (i.e. from J=2-1 to J=7-6)
datasets of interferometric CS observations for hot cores. CS data
presented in \citet{Haus95, Macd96, Beut02, Belt05} and
\citet{Leur07} focuss only on certain transitions of CS or on
isotopologues of CS. On the other hand, multiples lines of
$^{12}$C$^{32}$S are observed in galactic hot cores but only using
single-dish instruments. Despite these limitations, qualitative
comparison of our results with the data presented in
\citet{Mura94, Chan97} and more recently in \citet{Wu10} can be
made. Of particular interest, \citet{Wu10} found that the CS(2-1)
transition is indeed less compact than the CS(7-6) as seen in maps
of 50 massive very dense galactic sources. This agrees very well
with our model predictions and seems to confirm that the high-J CS
lines are one of the best tracers of very dense compact gas (i.e.
better than HCN). Similarly, they found that their mean and median
linewidth increase for high-J CS lines, which is also in agreement
with our model predictions. Here we do not attempt to model any of
the sources of their sample but the fact that our theoretical
approach and their observational results converge is encouraging.

Fig. \ref{fig:5} thus used by observers to estimate integration
times as it gives the expected flux of the first 15 transitions of
CS at various resolution (i.e. various radii). The fluxes are
 generally produced by a combination of a chemical model, line and
continuum radiative transfer code and telescope convolution
algorithm\footnote{We will publish on the web the deconvolved
fluxes
  as well as the convolved
fluxes. See http://www.homepages.ucl.ac.uk/$\sim$ucapdwi/interface/.}.
The deconvolved fluxes presented in Fig. \ref{fig:5} can be used
to estimate fluxes for other telescope parameters and
source distances. For different source types and masses, an observer
can scale these results, but the scaled results would be approximate
at best. With interferometers such as the IRAM-Plateau de Bure,
spatial resolutions up to $7\times 10^{16}$ cm are already
accessible (for a distance of 450 pc). As an example,
\citet{DeVi02} already performed a detailed study of HC$_{3}$N in
the Orion KL hot core. With Fig. \ref{fig:5}, we show that similar
studies are possible for CS. Indeed, a resolution of 5 $''$ is
reached in \citet{DeVi02} work, which corresponds to 0.01 pc (i.e.
3.856$\times 10^{16}$ cm) i.e. the CS UCC zone emission. More
information on the structure of the CS emission in hot cores will
be obtainable soon with ALMA.

\section{Conclusions}\label{sec:con}

We have performed a systematic theoretical study of the line
intensities and the properties of CS in archetypical hot core
environments. We have coupled via a user-friendly interface, a
large grid of chemical models with SMMOL radiative transfer code
and obtained line profiles of the CS(1-0) to CS(15-14) lines for a
variety of density, temperature, size and age. We also provide
(Fig. \ref{fig:5}) observers with estimates of line fluxes at
various resolution for the first 15 transitions of CS molecule.

Our main conclusions are:

\begin{itemize}
\item the CS fractional abundance is highest in the innermost
parts of the UCC whatever the age of the hot core. This confirms
the CS molecule as one of the best tracers of the very dense gas
component (see Sect. \ref{subsec:zone}).\\
\item the high-J CS lines have the strongest line fluxes, and the
linewidths are broader than those of low-J CS lines (see
Sect. \ref{subsec:stru}).\\
\item The peak antennae temperature of all the CS transitions
except for the CS(1-0) line is a good tracer of the kinetic
temperature inside the hot core because it
is very sensitive to its changes (see Sect. \ref{subsec:stru}).\\
\item In the Envelope, the older the hot core, the stronger the
self-absorption of CS. The best tracer of age seems to be the
CS(1-0) line which show the largest variations in fluxes with
respect to the time (see Sect. \ref{subsec:age}).\\
\item The CS(1-0) flux is coming mainly from the Envelope while
the high-J CS line fluxes are better tracers of the UCC zone
(see Sect. \ref{subsec:zone}).\\
\end{itemize}

\section*{Acknowledgments}

EB acknowledges financial support from STFC. Authors acknowledges
financial support from the Miracle Astrophysics High Performance
Computing Project.

\bibliographystyle{apj}
\bibliography{references}

\begin{landscape}
\begin{table}
\caption{Summary of model runs. ``-'' means no density or
temperature profiles whereas ``+'' means a profile has been
applied. The CS line fluxes are displayed in units of $10^{-15}$
Wm$^{-2}$. For each line, we specify the corresponding frequency
in GHz. \textbf{We have giving examples of paired models (see brackets
in columns 1 and 2) representing the coupling of the UCC with the
Envelope for a particular age. In column 9 we have listed only the
values of the CS abundance found at the outer and inner shells for each
model. The evolution of the CS abundance alongside the core
can thus be studied by looking continously at the columns 9 of the models
belonging to the same pair. }}\label{tab:models}
\begin{tabular}{llccccccccccc}
\hline &&Model & Type & size & density& temperature & age  & Frac.
abund. of & CS(1-0) & CS(3-2) & CS(7-6) & CS(10-9)\\
&&name  & & (pc) & profile & profile & (yrs) & CS (outer-inner
shell) & 48.99 GHz & 146.97 GHz & 342.88 GHz & 489.75 GHz\\
\hline
\multirow{2}{*}{\Big\{}& &HC$_{1}$ & UCC &0.03 & - & - & $3.2\times 10^{3}$ & $7.66\times 10^{-9}$ & $1.14 \times 10^{-4}$ & $1.85 \times 10^{-2}$ & $2.38 \times 10^{-1}$ & $3.90 \times 10^{-1}$\\
&&HC$_{2}$ & Env. &0.15 &  - & - & $3.2\times 10^{3}$ & $1.97\times 10^{-9}$ & $3.02 \times 10^{-3}$ & $4.03 \times 10^{-2}$ & $1.07 \times 10^{-1}$ & $1.31 \times 10^{-1}$\\
\multirow{2}{*}{\Big\{}&&HC$_{3}$ & UCC &0.03 &  + & - & $3.2\times 10^{3}$ & $1.44-1.72\times 10^{-8}$ &$1.32 \times 10^{-4}$ & $2.13 \times 10^{-2}$ & $2.35 \times 10^{-1}$ & $3.67 \times 10^{-1}$\\
&&HC$_{4}$ & Env. &0.15 &  + & - & $3.2\times 10^{3}$ & $3.77-2.34\times 10^{-9}$ & $2.37 \times 10^{-3}$ & $3.40 \times 10^{-2}$ & $7.40 \times 10^{-2}$ & $7.38 \times 10^{-2}$\\
\multirow{2}{*}{\Big\{}&&HC$_{5}$ & UCC &0.03 &  + & + & $3.2\times 10^{3}$ & $0.55-2.0\times 10^{-8}$ &$5.92 \times 10^{-5}$ & $9.81\times 10^{-3}$ & $1.21 \times 10^{-1}$ & $1.90 \times 10^{-1}$\\
&&HC$_{6}$ & Env. &0.15 & + & + & $3.2\times 10^{3}$ & $4.33-2.37\times 10^{-9}$ &$3.10 \times 10^{-3}$ & $2.89 \times 10^{-2}$  & $5.70 \times 10^{-2}$  & $5.90 \times 10^{-2}$\\
|&&HC$_{7}$ & UCC &0.03 & + & + & $1.0\times 10^{3}$ & $2.58-3.28\times 10^{-9}$ &$2.36 \times 10^{-5}$  & $4.39 \times 10^{-3}$ & $6.97 \times 10^{-2}$ & $1.04 \times 10^{-1}$\\
$\mid$&|&HC$_{8}$ & UCC &0.03 & + & + & $2.0\times 10^{3}$ & $3.76-9.82\times 10^{-9}$ & $3.56 \times 10^{-5}$  & $6.39 \times 10^{-3}$ & $9.22 \times 10^{-2}$ & $1.42 \times 10^{-1}$\\
$\mid$&$\mid$&HC$_{9}$ & UCC &0.03 & + & + & $5.0\times 10^{3}$ & $0.82-3.71\times 10^{-8}$ & $1.10 \times 10^{-4}$  & $1.52 \times 10^{-2}$  & $1.53 \times 10^{-1}$ & $2.42 \times 10^{-1}$\\
$\mid$&$\mid$&HC$_{10}$ & UCC &0.03 & + & + & $7.0\times 10^{3}$ & $1.16-5.89\times 10^{-8}$ &$1.73 \times 10^{-4}$  & $1.96 \times 10^{-2}$  & $1.73 \times 10^{-1}$ & $2.74 \times 10^{-1}$\\
$\mid$&$\mid$&HC$_{11}$ & UCC &0.03 & + & + & $1.0\times 10^{4}$ & $1.64-9.33\times 10^{-8}$ &$2.61 \times 10^{-4}$  & $2.35 \times 10^{-2}$ & $1.88 \times 10^{-1}$ & $3.00 \times 10^{-1}$\\
$\mid$&$\mid$&HC$_{12}$ & UCC &0.03 & + & + & $5.0\times 10^{4}$ & $0.55-4.92\times 10^{-7}$ &$7.59 \times 10^{-4}$  & $3.28 \times 10^{-2}$  & $2.29 \times 10^{-1}$ & $3.64 \times 10^{-1}$\\
$\mid$&$\mid$&HC$_{13}$ & UCC &0.03 & + & + & $1.0\times 10^{5}$ & $8.89-4.79\times 10^{-7}$ &$9.37 \times 10^{-4}$ & $3.48\times 10^{-2}$  & $2.39 \times 10^{-1}$  & $3.74 \times 10^{-1}$\\
$\mid$&$\mid$&HC$_{14}$ & Env. &0.15 & + & + & $1.0\times 10^{3}$ & $3.42-2.62\times 10^{-9}$ &$2.66 \times 10^{-3}$ & $2.79 \times 10^{-2}$ & $5.52 \times 10^{-2}$  & $5.60 \times 10^{-2}$\\
|&$\mid$&HC$_{15}$ & Env. &0.15 & + & + & $2.0\times 10^{3}$ & $3.81-2.44\times 10^{-9}$ &$2.86 \times 10^{-3}$ & $2.83 \times 10^{-2}$ & $5.59 \times 10^{-2}$  & $5.72 \times 10^{-2}$\\
&|&HC$_{16}$ & Env. &0.15 & + & + & $5.0\times 10^{3}$ & $5.15-2.50\times 10^{-9}$ &$ 3.47\times 10^{-3}$ & $2.97 \times 10^{-2}$  & $5.86 \times 10^{-2}$ & $6.17 \times 10^{-2}$\\
&&HC$_{17}$ & Env. &0.15 & + & + & $7.0\times 10^{3}$ & $6.16-2.83\times 10^{-9}$ &$3.87 \times 10^{-3}$ & $3.06 \times 10^{-2}$  & $6.04 \times 10^{-2}$ & $6.47 \times 10^{-2}$\\
&&HC$_{18}$ & Env. &0.15 & + & + & $1.0\times 10^{4}$ & $7.74-3.51\times 10^{-9}$ &$4.42 \times 10^{-3}$ & $3.17 \times 10^{-2}$ & $6.28 \times 10^{-2}$ & $6.85 \times 10^{-2}$\\
&&HC$_{19}$ & Env. &0.15 & + & + & $2.0\times 10^{4}$ & $13.77-7.29\times 10^{-9}$ &$5.83 \times 10^{-3}$  & $3.45 \times 10^{-2}$ & $6.88 \times 10^{-2}$ & $7.73 \times 10^{-2}$\\
&&HC$_{20}$ & Env. &0.15 & + & + & $5.0\times 10^{4}$ & $2.79-2.06\times 10^{-8}$ & $7.45 \times 10^{-3}$  & $3.82 \times 10^{-2}$ & $7.70 \times 10^{-2}$ & $8.78 \times 10^{-2}$\\
&&HC$_{21}$ & Env. &0.15 & + & + & $7.0\times 10^{4}$ & $3.45-2.58\times 10^{-8}$ & $7.90 \times 10^{-3}$  & $3.92 \times 10^{-2}$ & $7.94 \times 10^{-2}$ & $9.05 \times 10^{-2}$\\
&&HC$_{22}$ & Env. &0.15 & + & + & $1.0\times 10^{5}$ & $4.17-3.21\times 10^{-8}$ &$8.27 \times 10^{-3}$  & $4.03 \times 10^{-2}$ & $8.18 \times 10^{-2}$ & $9.31 \times 10^{-2}$\\
&&HC$_{23}$ & Env. &0.15 & + & + & $5.0\times 10^{5}$ & $9.53-7.29\times 10^{-8}$ &$9.79 \times 10^{-3}$  & $4.58 \times 10^{-2}$ & $9.43 \times 10^{-2}$ & $1.06 \times 10^{-1}$\\
&&HC$_{24}$ & Env. &0.15 & + & + & $1.0\times 10^{6}$ & $11.89-1.33\times 10^{-8}$ &$1.00 \times 10^{-2}$  & $4.61 \times 10^{-2}$ & $9.48 \times 10^{-2}$ & $1.05 \times 10^{-1}$\\
\hline
\end{tabular}
\end{table}
\end{landscape}

\end{document}